\documentclass[sigconf,natbib=false]{acmart}
\usepackage{lipsum}
\usepackage{listings}
\usepackage{float}
\restylefloat{table}
\AtBeginDocument{%
  }

\setcopyright{acmcopyright}
\copyrightyear{2023}
\acmYear{2023}
\acmISBN{979-8-4007-0148-1/23/06}
\acmDOI{10.1145/3587819.3592542}

\RequirePackage[
  datamodel=acmdatamodel,
  style=acmnumeric,
  ]{biblatex}

\begin{document}

\title{Everybody Compose: Deep Beats To Music}

\author{Conghao Shen}
\authornote{These authors contributed equally to this work.}
\affiliation{%
  \institution{Stanford University}
  \country{United States}
}
\email{tomshen@stanford.edu}

\author{Violet Z. Yao}
\authornotemark[1]
\affiliation{%
  \institution{Stanford University}
  \country{United States}
}
\email{vyao@stanford.edu}

\author{Yixin Liu}
\authornotemark[1]
\affiliation{%
  \institution{Stanford University}
  \country{United States}
}
\email{yixinliu@stanford.edu}

\acmYear{2023}\copyrightyear{2023}
\acmConference[MMSys '23]{Proceedings of the 14th ACM Multimedia Systems Conference}{June 7--10, 2023}{Vancouver, BC, Canada}
\acmBooktitle{Proceedings of the 14th ACM Multimedia Systems Conference (MMSys '23), June 7--10, 2023, Vancouver, BC, Canada}
\acmPrice{15.00}
\acmDOI{10.1145/3587819.3592542}
\acmISBN{979-8-4007-0148-1/23/06}

\renewcommand{\shortauthors}{C. Shen, V. Yao, Y. Liu}

\begin{abstract}
This project presents a deep learning approach to generate monophonic melodies based on input beats, allowing even amateurs to create their own music compositions. Three effective methods - LSTM with Full Attention, LSTM with Local Attention, and Transformer with Relative Position Representation - are proposed for this novel task, providing great variation, harmony, and structure in the generated music. This project allows anyone to compose their own music by tapping their keyboards or ``recoloring'' beat sequences from existing works.
\end{abstract}



\maketitle

\section{Introduction}
Artificial Intelligence has been widely applied to the domain of Arts. There have been many deep learning models that successfully generate paintings, music, stories, etc. It’s fascinating that a well-trained deep learning model could take a simple-format input and produce much richer content with higher aesthetic value. For example, SketchyGAN can synthesize realistic images from a simple sketch\cite{chen2018sketchygan}, helping novice painters to get a richer version of their original sketches. We realized that this can be applied to the area of music generation. Beats are one of the essential components of music and it is relatively simple compared to chords, harmony, and melody. Without any expert music knowledge, everyone could describe the beats simply by clapping their hands. The motivation of this project is to use deep learning model to generate monophonic melodies that correspond to the input beats. This project allows even amateurs to create their own piece of music from simple beats, allowing everyone to enjoy the satisfaction of music composition. Furthermore, professional composers could also use this model to get inspiration during their music production. Our implementation is available on  \url{https://github.com/tsunrise/everybody-compose}.
\section{Related Work and Novelty}
There have been various approaches to generating music using deep learning models. Some existing approaches use autoregressive Recurrent Neural Networks to generate the sequence of output music. LSTM has been applied to music generation tasks to solve the vanishing gradients problem in simple RNN models and to learn the complex relationship between chords\cite{eck2002first}\cite{choi2016text}. Performance RNN is an LSTM-based model that uses event-based representation as its output\cite{performance-rnn-2017}. Even though Performance RNN generates music that sounds plausible for a short while, it lacks long-term structure and coherence. Lookback RNN and Attention RNN proposed from the Magenta Project obtains the ability to learn the long-term structure of music by inputting events 1 and 2 bars ago or looking at the output from the last few steps when generating output for the current step\cite{waite_2016}.

Transformers have also been used to capture the long-term structure of music generation. MuseNet\cite{payne_2019} uses a large-scale transformer model to predict the next token in the music sequence, and it is also able to take instrument and composer as input for music generation. Music Transformer\cite{huang2018music} uses relative attention to focus on relational features which outperform the original transformer model.

Generative Adversarial Networks (GAN)\cite{goodfellow2020generative} have been applied to several domains including music generation. Transformer-GAN\cite{transformer-gan} uses Transformer-XL as the generator and BERT as the discriminator.  It achieves better performance compared to transformer models that maximize likelihood alone. 

However, we want to highlight that none of the previous approaches takes beats as inputs. Instead of sampling based on \\ $p(y_i|y_1,y_2,...,y_{i-1})$, our model introduces a hybrid approach, allowing both teach forcing from labels $\mathbf{y}$ and user guidance $\mathbf{x}$ so that the model samples notes according to $p(y_i|y_1,...y_{i-1},x_1,...,x_i)$.

\section{Dataset and Features}
\begin{figure}[H]
    \centering
    \vspace*{-0.17in}
    \includegraphics[width=0.4\textwidth]{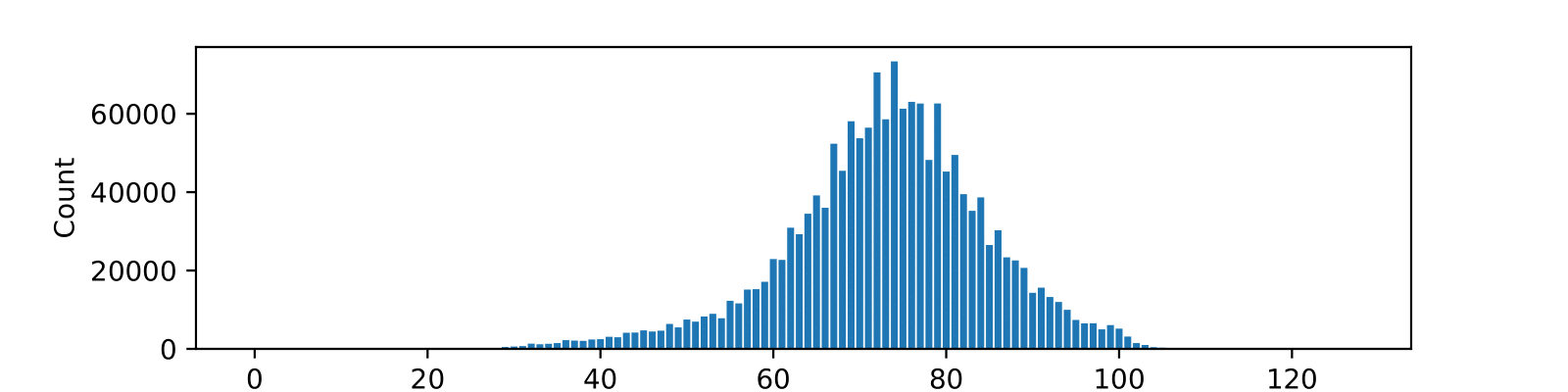}
    \caption{Distribution of note pitches in MAESTRO Dataset}
    \label{fig:maestro}
    \vspace*{-0.1in}
\end{figure}

Our project uses the MAESTRO dataset \cite{maestro}, a large-scale collection of piano performances MIDI files compiled by the Magenta research team at Google. The dataset consists of over 200 hours of recordings and 6.18 million notes from the competition virtuoso pianists in the International Piano-e-Competition \cite{PianoECompetition}. By using this dataset, we can train models to generate melodies of reasonable quality. A potential limitation is that this dataset only covers the Classical genre, making our model not able to generate melodies of other genres like Pop effectively, but inside the Classical genre, the dataset contains many musical styles, which is enough for our model to generalize well. \ref{fig:maestro} shows the distribution of note pitches in the dataset. The note pitches follow the normal distribution with the mean around 78, which resembles the note pitch distribution of a typical classical piano performance.

\textbf{Data Preprocessing}. A MIDI file consists of a sequence of MIDI events, and in piano performance, each event can be one of the four types: \texttt{NOTE\_START}, \texttt{NOTE\_END}, \texttt{REST}, \texttt{VELOCITY\_SHIFT}. The first two control the position and relative length of notes, and the last two controls the timing. In prior work such as \cite{performance-rnn-2017}, and \cite{huang2018music}, the sequence model learns and generates those events directly, but our model uses a novel and simpler representation. We first use the \texttt{note-seq} library \cite{magenta-note-seq} to convert the MIDI events to an overlapping interval of notes. Since our current model supports only monophonic melody, we have applied a melody inference algorithm in \cite{magenta-note-seq} to make those intervals disjoint without losing the overall musical structure. This algorithm divides the continuous time space into frames, and for each frame, computes the possible melody event with the highest likelihood, and then uses Viterbi Algorithm to compute the most likely sequence of melody events using the likelihood computed. In the case of chords, the algorithm favors the highest note in the chord. 

\textbf{Features Representation}. The next step is to convert the disjoint time interval to a sequence. Our model represents the piano performance as a sequence of disjoint notes. The feature $X$ is a sequence of ``beats'' where $X^{\langle t \rangle}$ is a tuple such that $X^{\langle t \rangle}_0$ is the rest time after the release of previous note at timestep $t-1$, and $X^{\langle t \rangle}_1$ is the duration of current note at time $t$. The label $y$ is a sequence of note pitches, ranging from 0 to 127, where $y^{\langle t \rangle}$ corresponds to the note pitch at time $t$ whose beat is $X^{\langle t \rangle}$. Since no performance uses the pitch 0, we use 0 to represent the start of a sequence $y^{\langle 0 \rangle}$. Doing so, for each sample, the feature has shape \texttt{(sequence length, 2)}, and the label has shape \texttt{(sequence length,)}.

\textbf{Optimization}. A typical music performance can have more than 2000 disjoint notes, but many sequence models for NLP cannot handle such long sequences. For example, RNN and even LSTMs can suffer vanishing gradient problem when the sequence length is greater than 128, and transformers will take an extremely long time to train and infer when a sequence becomes long because its runtime is quadratic to the sequence length \cite{NIPS2017_3f5ee243}. To overcome this issue, we implemented random slicing in our \texttt{DataLoader}, wherein for each epoch, we randomly take a slice of fixed length for each sample. This method helps our model to converge faster without losing generality. 
In addition, we realized that preprocessing takes a significant time during the training process - it takes around 3 hours on AWS \texttt{m2.xlarge} instance. To alleviate this problem, we host the preprocessed data on Cloudflare, so now everyone takes only 5 seconds to download and can directly start training then.

\textbf{User Input and ``Recoloring''}. We have written a sampling utility that allows users to write beats by tapping their keyboard. Then, the beats will be converted to beats sequence and fed to our sampling algorithms for note inference. The user will then get a MIDI file to hear the generated melody. The sampling utility also supports ``recoloring'', where it takes a sample from the dataset, extra its beats, and inference a new melody from the beats. 
\section{Methods}

\subsection{Baseline: Decoder Only Vanilla RNN}
We implemented an autoregressive decoder-only vanilla RNN model as our baseline. It takes input beats Sequence $X$ and outputs notes sequence $Y$ with the same length. For every time stamp $t$, we concatenate the input $x^{\langle t \rangle}$ with the note embedding $y^{\langle t-1 \rangle}$ to allow teacher forcing. Then the concatenation result is fed through a dense layer before feeding into the two-layer single-directional recurrent neural network. Then the output of the recurrent neural network is forwarded through another dense layer with softmax activation to output notes prediction. We also add a residual connection between the output of concatenation and before the RNN layers allowing an alternative information flow path. The baseline model suffers from the vanishing gradient problem. Also, the model can't access future beats information, and it is too simple to capture the long-term dependencies in the music.

\subsection{LSTM with Full Attention}
We implemented an LSTM model with Full Attention which is similar to the architecture in \cite{bahdanau2014neural}. To mitigate the vanishing gradient problem in the baseline model, we replaced the vanilla RNN cell with LSTM cell, which introduces separate cell states to encode long-term memory and extra gates to decide how much past information is kept\cite{hochreiter1997long}. The attention mechanism is added to allow the decoder to utilize the most relevant information in the encoder output. 

The model uses a pre-attention bidirectional LSTM and a post-attention single-directional LSTM. The pre-atttention bidirectional LSTM takes the input of beats sequence $X$, and outputs a sequence of annotations ($h^{\langle 1 \rangle}$, $h^{\langle 2 \rangle}$, … $h^{\langle T \rangle}$). The context vector $context^{\langle t \rangle}$ is computed as a weighted sum of annotations: $context^{\langle t \rangle} = \sum_{j=1}^{T}\alpha_{tj}h^{\langle j \rangle}$. The attention weight $\alpha_{tj}$ is computed with the previous hidden state of post-attention LSTM and annotations through a one-hidden layer neural network. Then the context vector at the current timestamp is concatenated with the note embedding in the previous timestamp to feed into the post-attention LSTM to obtain the output notes. One downside of the model is the cost of training is quadratic due to the introduction of the attention mechanism.

\subsection{LSTM with Local Attention}
We have designed a model architecture called LSTM with Local Attention that performs better than the attention model and is much easier to train. Our architecture contains a bidirectional LSTM as the first layer and a single-directional LSTM as the second layer. The bidirectional accepts input sequence $X$ and outputs the final hidden state and context sequence $h$ where the $c^{\langle t \rangle}$ is a concatenation of the hidden state of forward direction and backward direction at timestep $t$. The standard LSTM is an autoregressive model where the input at timestep $t$ is a concatenation of previous note $y^{\langle t-1 \rangle}$ and hidden state $h^{\langle t \rangle}$. The critical difference is that the attention at timestep $t$ is local - instead of taking a weighted average of the entire context sequence $\sum_{j=1}^{T} \alpha_{tj} h^{\langle j\rangle}$, the second LSTM only takes $h^{\langle t\rangle}$. This idea works because of the unique properties in our problem settings - the input beats and output notes sequence have the same length, and unlike machine translation where the target word may correspond to different positions in the source word, beats and notes have a strong one-to-one relationship, so using the context vector at the same timestep gives enough information to infer the next note. This model is also inspired by the encoder-decoder, where the initial state of the second LSTM is set to be the final hidden state of the first bidirectional LSTM, allowing the decoder to have a better awareness of the overall beat structures. 

\subsection{Transformer with Relative Position Representation}
Transformers avoid the dependence on the recurrence architecture and utilize self-attention to allow global dependencies between inputs and outputs. Our Transformer model utilizes an encoder-decoder architecture. The encoder takes in a sequence of beats as input, projects them into a dense vector, and feeds the vectors into a self-attention sub-layer and a feedforward sub-layer. Each sub-layer is followed by a residual connection and layer normalization to facilitate information propagation back into deeper layers. The decoder is autoregressive, taking in notes predicted so far, projecting them into embeddings, and then passing the vector to a self-attention sub-layer, an encoder-decoder attention sub-layer for reference to the encoded state, and a feedforward sub-layer. During training, an input mask is used by the decoder to prevent it from accessing future inputs. Additionally, each sub-layer is followed by a residual connection and layer normalization to improve gradient flow. Finally, a generator linear layer decodes the feedforward output to the space of possible notes. To increase the representation power of the network, multiple encoder and decoder layers are stacked.

\begin{equation} \label{relattention}
RelativeAttention(Q, K, V) = Softmax(\frac{QK^T + QE^{rT}}{\sqrt{D}})V
\end{equation}

To represent the sequence order, Transformers add sinusoidal positional encodings to their inputs, aiding the models in learning the absolute position of each input element. While absolute position representation helps learn the global timing and pitch of a melody, relative distances are also valuable in capturing pairwise relationships between input elements. Music Transformer \cite{huang2018music} applies the idea of relative self-attention \cite{shaw2018self} in the music generation space, reducing the memory requirements from $O(T^2D)$ to $O(TD)$. Thus, inspired by the success of Music Transformer\cite{huang2018music} and LSTM with Local Attention, we implement relative position representation to facilitate learning pairwise relationships. In this algorithm, a separate relative position embedding $E^r$ of shape $num\_heads \times T \times embed\_dim$ is learned for each possible pairwise distance, separately for each attention head. In Equation \ref{relattention}, Query, Key, and Value matrices are denoted as Q, K, and V, respectively. An additional $QE^{rT}$ of shape $T \times T$ is added in calculating attention weights. We extend the implementation of Music Transformer, which models MIDI events and employs a decoder-only architecture, to an encoder-decoder architecture with relative attention in both modules to handle beats sequence as inputs and notes sequence as outputs. 



\subsection{Sampling and Searching}
\textbf{The State Machine Philosophy}. We have defined an interface for sampling, where it abstracts each model as a \textit{state machine} such that in each time step, it takes the current state and the previous sampled note, and outputs the next state and the probability distribution of the next note. It also has access to constants that do not change during the sampling or beam search. In LSTM, the state contains the hidden state of the LSTM cell and the current position, and constants contain the context sequence. In the transformer model, the state contains only the current position, and constants contain the encoder memory. This state machine model helps us to apply a search algorithm to different models without code duplication and allows us to keep track of states in beam search easily.

\begin{table*}[!htbp]
\centering\small
\begin{tabular}{||c c c||} 
 \hline
Model & Train Accuracy & Validation Accuracy  \\ 
 \hline\hline
Baseline (Decoder Only Vanilla RNN) & 0.4170 & 0.3908 \\ 
 \hline
LSTM with Full Attention & 0.5506 & 0.4218 \\
 \hline
LSTM with Local Attention & \textbf{0.7079} & \textbf{0.5077} \\
 \hline
Transformer & 0.3897 & 0.3797 \\
 \hline
Transformer with Relative Position Representation & 0.5034 & 0.4690 \\
 \hline
\end{tabular}
\caption{\label{all_results} DeepBeats Models Train/Validation Accuracy}
\end{table*}

\textbf{Stochastic Search and Heuristics}. Randomness have played an important factor in increasing the quality of generated melodies. In some prior work like \cite{performance-rnn-2017}, \cite{transformer-gan}, randomness has helped boosted the variability of the generated artifacts and balanced exploration and exploitation during the search. In our work, we applied randomness in our search, where in each timestep, we query the state machine, get the distribution of the next note, and randomly select a note according to the queried distribution. We call this process \textit{stochastic search}. We have used several heuristics to ensure the quality of the generated notes while keeping the added creativity from randomness. We have used \textit{top-p sampling}, where we only consider the top $p$ proportion of probability mass for sampling, and \textit{top-k sampling}, where we only consider the top $k$ choices. We also used \textit{temperature} $T$ to adjust probability such that $\forall r: \tilde{P}(y^{(t+1)}=r|\mathbf{x}, \mathbf{y}) \propto P(y^{(t+1)}=r|\mathbf{x}, \mathbf{y})^{\frac{1}{T}}$. Doing so, $T>1$ gives more confidence to the note with larger likelihood, reducing the variability, and $T<1$ makes the distribution more uniform, increasing the variability. We also designed a heuristic called \textit{repeat decay} $\gamma$, where we reduce the likelihood of repeating the previous note by $\gamma$. That is: $\forall r: \tilde{P}(y^{(t+1)}=r|y^{(t)} = r, \mathbf{x}, \mathbf{y}) = (1-\gamma)P(y^{(t+1)}=r|\mathbf{x}, \mathbf{y})$. Doing so, we upper bound the probability of repeating the same note $N$ times by a constant $(1-\gamma)^{N-1}$, which decreases exponentially in $N$, making the generated melodies less repetitive and more interesting. In addition, we allow users to fix a few notes at the beginning as a hint. 

\textbf{Hybrid Beam Search}. To better balance creativity and the objective of maximizing sequence likelihood, we have combined the idea of stochastic search and beam search. In detail, suppose we have $N$ beams. There are two modes -  \textit{beam mode} and \textit{stochastic mode} for selecting the next $N$ beams. The \textit{beam mode} is the same as the original beam search, where for each beam, we query the model state machine and get the state and the conditional distribution for the next beam, and among the $N^2$ beams, calculate the likelihood for each corresponding sequence by summing the log of conditional likelihood, and select the top $N$. In \textit{stochastic} mode, for each beam, we sample the next note as the next beam and take the adjusted conditional likelihood according to the sampling heuristics. For each time-step, the sampler chooses \textit{beam mode} and \textit{stochastic mode} randomly according to a hyperparameter $p$ where $p$ is the probability of choosing beam mode. 

\section{Results and Discussion}


We perform hyperparameter tuning on various parameters such as learning rate, embedding dimension, hidden state dimension, and the number of encoder/decoder layers. The train/validation accuracy for the best-performing configuration for each of our methods is reported in Table \ref{all_results}. We establish a competitive baseline with a 39.08\% validation accuracy. LSTM with Full Attention achieves a 3.08\% improvement over the baseline, while the local attention mechanism brings a significant 8.59\% improvement. The vanilla Transformer achieves a modest 37.97\% validation accuracy, while the addition of relative position representation yields an 8.92\% improvement, illustrating the importance of learning relative pairwise relationships in note generation. We observe that Transformer models take longer to converge, and often require deeper stacked layers than those of LSTMs to reach comparable performance, resulting in more intensive computing costs.

For qualitative analysis, we use both beats in the dataset and user-inputted beats to generate notes and utilize \lstinline{SuperCollider} to play the midi notes generated by the models. We then examine the music quality in terms of the variety of notes used, the harmony of their sequence, and their smoothness. Also, musical composition heavily relies upon local and long-range context to construct periodicity and structures at different time scales. Our baseline model tends to generate simpler melodies and lacks long-range coherence due to the vanishing gradient problem and the inability to see the future beats. The LSTM with Full Attention model is able to generate smoother and more plausible melodies compared to the baseline model. A piece of melody may recur in the generated sequence, showing an improvement in long-term coherence. While it may continuously generate sequences of descending or ascending notes when the input beat sequence is long. The LSTM with Local Attention model achieves the best overall musical quality, with rich musical elements and great harmony, while it lacks long-term patterns because the memory capacity in the decoder of LSTM is bounded by the number of hidden states which can be limited. The Transformer with Relative Position Representation model better captures long-range coherence but tends to capture less local variation than the LSTM with Local Attention model, which suggests that the recurrence architecture empowers a better understanding of the local context. Sample generated melodies are available here \footnote{\url{https://tinyurl.com/everybodycompose}}.

\section{Conclusion and Future Work}
This study proposes three effective methods - LSTM with Full Attention, LSTM with Local Attention, and Transformer with Relative Position Representation - for the novel task of translating simple beats to music with great variation, harmony, and structure. We enable everybody, including amateurs and musicians, to compose their own music by tapping their keyboards or ``recoloring" beat sequences from existing works. Since music quality is subjective, for future work, we plan to conduct a larger-scale user study to gather feedback from both novices and professionals in order to iterate our models. To further increase the variation and diversity of generated music, we aim to extend our model output space from notes to chords.


\printbibliography

\appendix









\end{document}